\documentclass[conference]{IEEEtran}
\IEEEoverridecommandlockouts

\usepackage{cite}
\usepackage{amsmath,amssymb,amsfonts}
\usepackage{algorithmic}
\usepackage{graphicx}
\usepackage{textcomp}
\usepackage{xcolor}
\usepackage{subfig}
\usepackage{float}
\usepackage{multirow}
\usepackage{url}
\usepackage{textcase}
\def\BibTeX{{\rm B\kern-.05em{\sc i\kern-.025em b}\kern-.08em
    T\kern-.1667em\lower.7ex\hbox{E}\kern-.125emX}}
\usepackage{balance}

\newcommand{\etal}{\emph{et al.}}

\begin{document}

\title{Multi-resolution Encoding for HTTP Adaptive Streaming using VVenC}

\author{
\IEEEauthorblockN{Kamran Qureshi\textsuperscript{1},
Hadi Amirpour\textsuperscript{1}, and Christian Timmerer\textsuperscript{1} }\vspace{0.2cm}

\IEEEauthorblockA{\textsuperscript{1} Christian Doppler Laboratory ATHENA, Alpen-Adria-Universitat, Klagenfurt, Austria}

}

\maketitle

\begin{abstract}
HTTP Adaptive Streaming (HAS) is a widely adopted method for delivering video content over the Internet, requiring each video to be encoded at multiple bitrates and resolution pairs, known as representations, to adapt to various network conditions and device capabilities. This multi-bitrate encoding introduces significant challenges due to the computational and time-intensive nature of encoding multiple representations. Conventional approaches often encode these videos independently without leveraging similarities between different representations of the same input video. 
This paper proposes an accelerated multi-resolution encoding strategy that utilizes representations of lower resolutions as references to speed up the encoding of higher resolutions when using Versatile Video Coding (VVC); specifically in VVenC, an optimized open-source software implementation. For multi-resolution encoding, a mid-bitrate representation serves as the reference, allowing interpolated encoded partition data to efficiently guide the partitioning process in higher resolutions. The proposed approach uses shared encoding information to reduce redundant calculations, optimizing partitioning decisions.
Experimental results demonstrate that the proposed technique achieves a reduction of up to 17\% compared to medium preset in encoding time across videos of varying complexities with minimal BDBR/BDT of 0.12 compared to the fast preset. 

\end{abstract}

\begin{IEEEkeywords}
Compression, HTTP adaptive streaming, multi-resolution, multi-rate.
\end{IEEEkeywords}

\section{Introduction}

The rapid growth of streaming services has made efficient video delivery a fundamental requirement for modern content distribution. Among the various streaming techniques, HTTP Adaptive Streaming (HAS) has emerged as the industry standard due to its ability to dynamically adjust video quality in response to varying network conditions. To ensure smooth playback and adaptability, each video must be encoded at multiple resolutions and bitrates, known as representations~\cite{amirpour_deepstream_2022}, enabling seamless switching based on network bandwidth and device capability. However, encoding these multiple representations introduces significant computational challenges, as each resolution and bitrate combination typically requires separate encoding processes.

To meet the demands of large-scale, real-time streaming, efficient encoders have been developed to reduce the computational burden while maintaining high-quality compression. One of the most advanced solutions in this context is Versatile Video Coding (VVC)~\cite{bross_overview_2021}, which offers substantial improvements in compression efficiency -- up to 40\% compared to its predecessor, High-Efficiency Video Coding (HEVC)~\cite{sullivan_overview_2012} -- while supporting new media formats such as ultra-high-definition (UHD) and 360-degree/immersive video. However, the benefits of VVC come at a cost: its encoding process is computationally intensive, making real-time applications more challenging.

To mitigate VVC's complexity and enhance its suitability for practical streaming, an optimized open-source implementation known as VVenC~\cite{wieckowski_vvenc_2021} was developed. VVenC accelerates the encoding process by incorporating a range of speed-up techniques, such as early termination of encoding modes and optimized partitioning decisions, without sacrificing the compression efficiency inherent to VVC. As a result, VVenC enables faster encoding, making it a more feasible solution for real-time streaming scenarios while still leveraging the advanced compression capabilities of the VVC standard.

\begin{figure}
    \centering
    \includegraphics[width=1\linewidth]{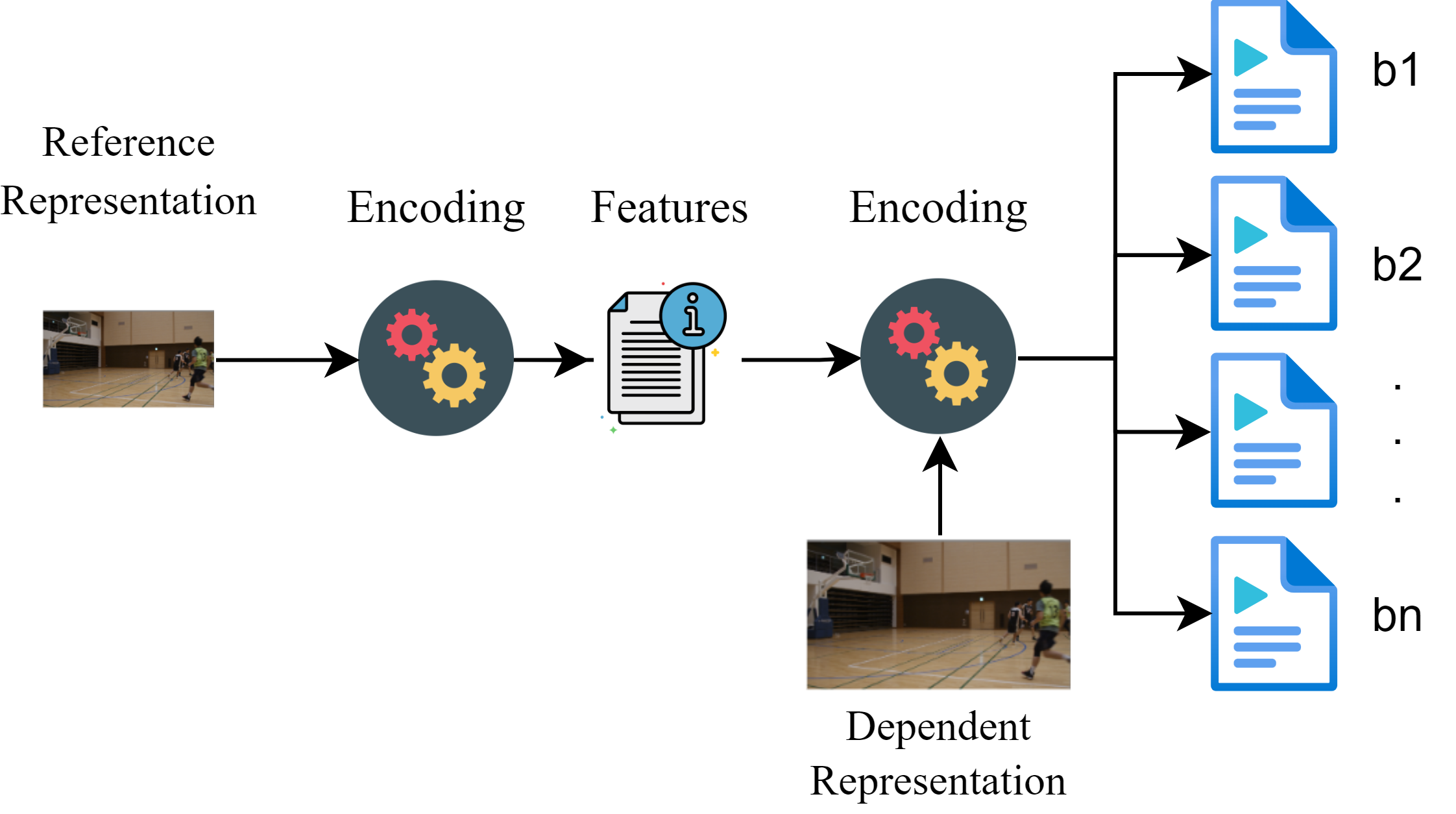}
    \caption{Overview of the MEVHAS framework. A reference representation is first encoded at a lower resolution, which helps accelerate the encoding of dependent representations at higher resolutions across all bitrates. }
    \label{multiresolution}
\end{figure}

Despite advances in compression efficiency with VVenC, encoding multiple representations for HAS remains a significant challenge, especially for live-streaming scenarios where time constraints are critical. 
This requires accelerated generation of multiple representations to meet time constraints. 

In recent years, substantial research has been dedicated to optimizing video encoding to reduce computational complexity and improve encoding speed, particularly for multiple-quality representations. Schroeder~\etal~\cite{schroeder2015multiC} introduced a method for high-resolution encoding that determines a threshold for early termination of low-resolution video representation encoding, thereby streamlining the process. In follow-up work, Schroeder~\etal~\cite{schroeder2015block} utilized the highest quality representation (QP 22) to accelerate the encoding of lower quality versions of the same video and later combined these strategies to identify specific encoding parameters that enhance the efficiency of dependent representations~\cite{schroeder2016efficient}. Meanwhile, Amirpour~\etal~\cite{amirpour_fast_2020} proposed a double bond approach, which predicts block partitioning for lower quality representations based on the highest quality version, facilitating better control of the intermediate quality representation. In further studies, Amirpour~\etal~\cite{amirpour2021towards} demonstrated that using a middle-quality representation as a reference for all other dependent representations effectively minimizes encoder complexity. Cetinkaya~\etal~\cite{cetinkaya2020fame} leveraged machine learning to expedite the encoding of multiple representations, focusing initially on parallel encoding. They later extended this work to incorporate multi-resolution encoding through machine learning models, which improved the encoding process further~\cite{cetinkaya2021fast}. Menon~\etal~\cite{menon2023emes} explored serial and parallel encoding optimizations, proposing multi-encoding schemes considering the trade-offs between compression efficiency and encoding time. In a different approach, Liu~\etal~\cite{liu2023preparing} applied a multi-rate encoding technique in VVenC, using lower-quality representations to save time when encoding higher-quality versions. 

Despite substantial efforts to reduce the encoding complexity of VVC, most optimizations have focused on VTM rather than VVenC, primarily targeting enhancements in intra-frame \cite{li2023texture, shang2023fast, yang2024fast, jiang2023low, song2024fast, chen2023speed, kherchouche2024rd, li2024fastIntra} and inter-frame \cite{kuang2022unified, li2024fast, liu2022light, peng2023classification, lin2024efficient} coding techniques. Moreover, while the multi-rate approach has been explored for VVC~\cite{liu2023preparing}, multi-resolution acceleration remains largely uncharted.

This research aims to achieve the speed-up by using the features from a lower resolution representation, also known as reference representation,  to accelerate the encoding of a higher resolution, also known as the dependent representation, across all required bitrates (b) in VVenC, as shown in Fig.~\ref{multiresolution}. This approach is called the \textbf{M}ultiresolution \textbf{E}ncoding in \textbf{V}VenC
for \textbf{H}TTP \textbf{A}daptive \textbf{S}treaming (\textbf{MEVHAS}) and is effective in streaming services where it can assist in adaptive bitrate streaming by providing more representations of the same content or can benefit real-time encoding, like live-streaming or video conferencing, to produce content at a faster rate.


\begin{figure}
    \centering
    \includegraphics[width=0.8\linewidth]{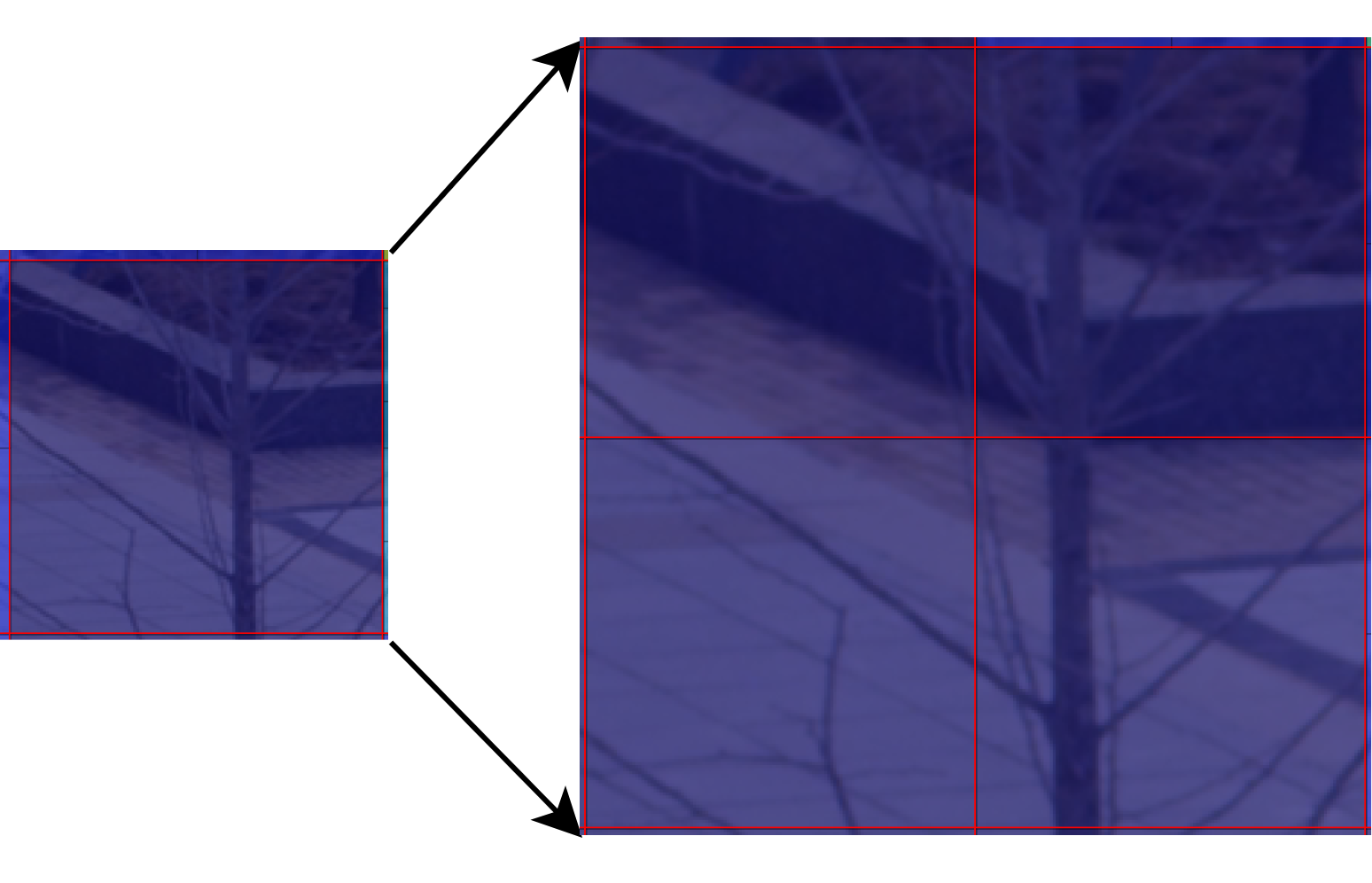}
    \caption{Low-resolution to high-resolution interpolation of one CTU.}
    \label{lowHighCTUInterpolation}
\end{figure}

\begin{figure}
    \centering
    \includegraphics[width=0.8\linewidth]{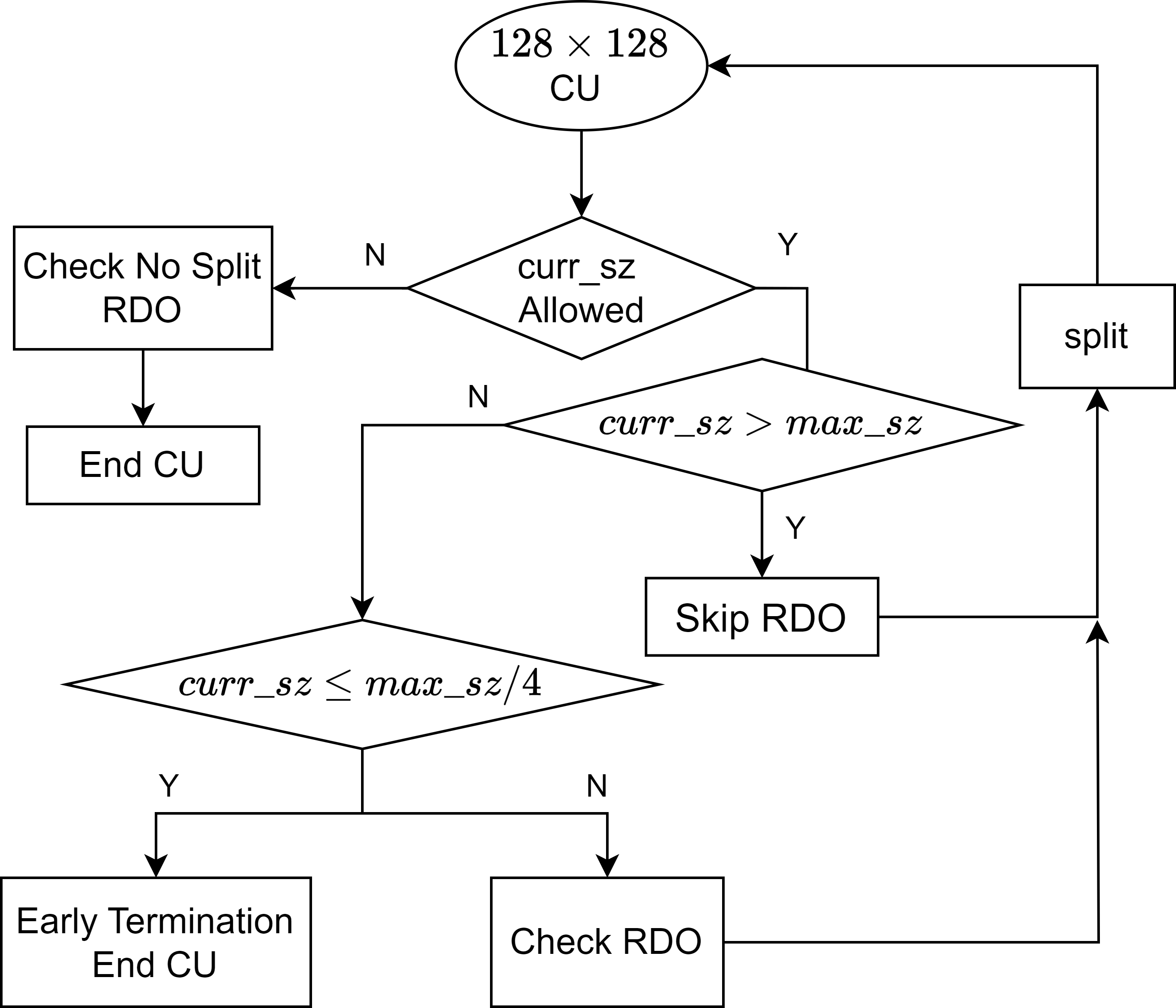}
    \caption{MEVHAS flow chart for encoding time reduction in VVenC.}
    \label{flowDiagram}
\vspace{-1em}
\end{figure}

\section{VVC Background}
VVC is a block-based hybrid video codec designed similarly to its predecessors. 
It delivers around 40\% bitrate reduction by supporting multiple new high-level extensions and block-level coding tools. 
The partitioning scheme integrated in VVC provides the most efficiency gains~\cite{franccois2020vvc} but at the expense of a significant increase in complexity~\cite{saldanha2020complexity,tissier2019complexity}. In the partitioning scheme of VVC, a frame is divided into fixed size blocks of $128\times128$ pixels referred to as Coding Tree Unit (CTU). Each CTU is processed individually and may or may not split into further blocks called Coding Units (CU). 
The split type is called no split (NS) if the block chooses not to divide. 
The splitting process is recursive, where a block can divide into four square sub-regions called a quad-tree (QT) split or can split into two rectangular sub-regions called binary tree (BT) split. It can also be divided into three rectangular sub-regions, with one region having twice the size of the other two regions. This type of split is called a ternary tree (TT) split. The partitioning scheme of VVC is prohibited from having further QT splits within a BT or TT split due to the cost of signaling overhead~\cite{huang2021block}, but it is allowed to further split into more BTs or TTs. 

Similarly to its predecessors, a CU in VVC is associated with its corresponding Prediction Units (PU) and Transform Units (TU). The concept of CU, PU, and TU is simplified by having PU and TU of the same sizes as the CU. The CU or TU is divided into subblocks only as an exception when subblock-based temporal motion vector prediction (SbTMVP)~\cite{yang2021subblock}, decoder-side motion vector refinement (DMVR)~\cite{gao2020decoder}, Intra subpartition (ISP)~\cite{de2019intra}, or subblock transform (SBT)~\cite{zhao2021transform} is performed.


\section{Multi-Resolution Encoding Approach}
The proposed method, "MEVHAS" uses low-resolution videos as reference representation for encoding high-resolution representations of the same videos. Specifically, low-resolution videos encoded at a QP of 37 are used as references for high-resolution encodings at QPs 42, 37, and 32, while low-resolution videos at QP 32 serve as references for high-resolution videos at QP 27. This choice is based on the observation that partitioning patterns from higher QPs (lower quality) do not effectively predict the partitioning needed for lower QPs (higher quality), as higher QPs explore fewer levels of detail in the partitioning hierarchy. By selecting QP 37 and QP 32 from low-resolution videos, we aim to utilize their structural similarities to enhance partitioning efficiency and accuracy during high-resolution encoding.

\begin{figure}
    \centering
    \includegraphics[width=0.9\linewidth]{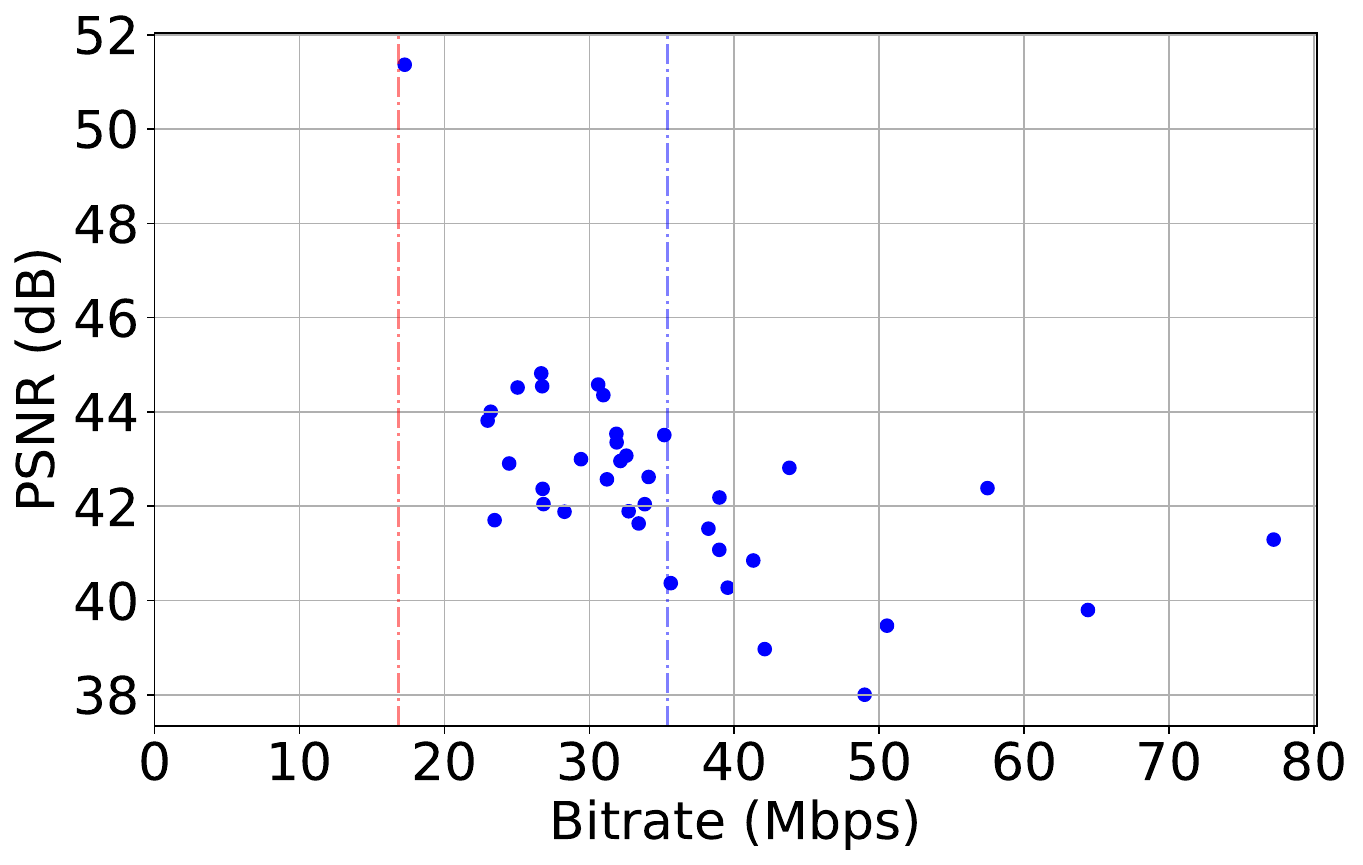}
    \caption{Bitrates of various Inter4K videos at QP 22 exceeding on average 16.8 Mbps.}
    \label{QP22bitrate}
\end{figure}

Apple's HTTP Live Streaming (HLS) has capped its live streaming bitrate at 16.8 Mpbs for 4K videos \cite{apple_http_2015} as higher bitrates result in near-lossless or lossless encoding. Such high bitrate levels are unsuitable for live streaming, where the goal is to maintain a balance between compression quality and data rate to avoid network strain and ensure efficient transmission. QP values lower than 27 exceed this bitrate. 
This effect can clearly be seen in Fig. \ref{QP22bitrate}, where the bitrate on average (blue line) of all the videos when encoded with QP 22 is higher than 16.8 Mbps (red line).
Therefore, QPs from 42 to 27 are chosen to sustain practical compression efficiency while preventing excessive bitrate increases.


MEVHAS consists of two stages: \textit{(i)} partitioning information is extracted from low-resolution video frames and subsequently interpolated; \textit{(ii)} the interpolated partitioning patterns are then utilized as an informed estimate to determine optimal partitioning for high-resolution video content. 

Fig.~\ref{lowHighCTUInterpolation} illustrates the width and height of a Coding Tree Unit (CTU) at a lower resolution. After interpolation, the partitioning of a single lower-resolution CTU corresponds to four CTUs at a higher resolution. Each block partitioned with a red line is one CTU of higher resolution. The data is first structured as an $8\times8$ matrix, representing the widths and heights of each Coding Unit (CU) within a CTU for low-resolution. 
The size of the matrix is doubled, and the values are scaled by a factor of two to create an estimate for the partitioning patterns of higher-resolution content. 
The modified matrix is flattened to produce two rows per CTU. The first row represents 256 width values, and the second row represents 256 height values for one $128\times128$ CTU.  
The width and height of each interpolated CU are used to calculate max\_sz, which denotes the area of the interpolated CU.

Fig.~\ref{flowDiagram} shows the flow of the algorithm applied in the second stage. The encoder loop starts with a $128\times128$ CU and recursively partitions it to the minimum allowed size ($8\times8$). For each recursive call, the area of the current partition is calculated as curr\_sz. An exception is handled for curr\_sz, where if the CU cannot further split into a quad-tree based on the encoder limitation, a default rate-distortion-optimization (RDO) check is allowed. In case of no restriction from the encoder, the curr\_sz is checked against the max\_sz. Above the max\_sz all RDOs are skipped, but the recursive partitioning is allowed. In the event that curr\_sz is less than or equal to max\_sz, the curr\_sz is checked against the minimum depth allowance. 
The minimum depth allowance is determined based on the max\_sz parameter, enabling the CU to extend two partitioning depths below the initial depth achieved through the estimated partitioning pattern. This approach is based on the observation that the available partitioning information originates from low-resolution video encoded at a lower bitrate.
When encoding at lower bitrates, the encoder selects higher partition depths to optimize the trade-off between computational complexity over compression efficiency. In contrast, the encoder may prefer lower partition depths to enhance compression efficiency and maintain higher quality at higher bitrates. All the RDO and recursive partitioning are skipped below $max\_sz/4$. The whole process is repeated for each high-resolution CU and thus adapts the skipping of RDO and recursive partitioning dynamically for each CU based on the area of interpolated CU.



\begin{figure}
    \centering
    \includegraphics[width=0.79\linewidth]{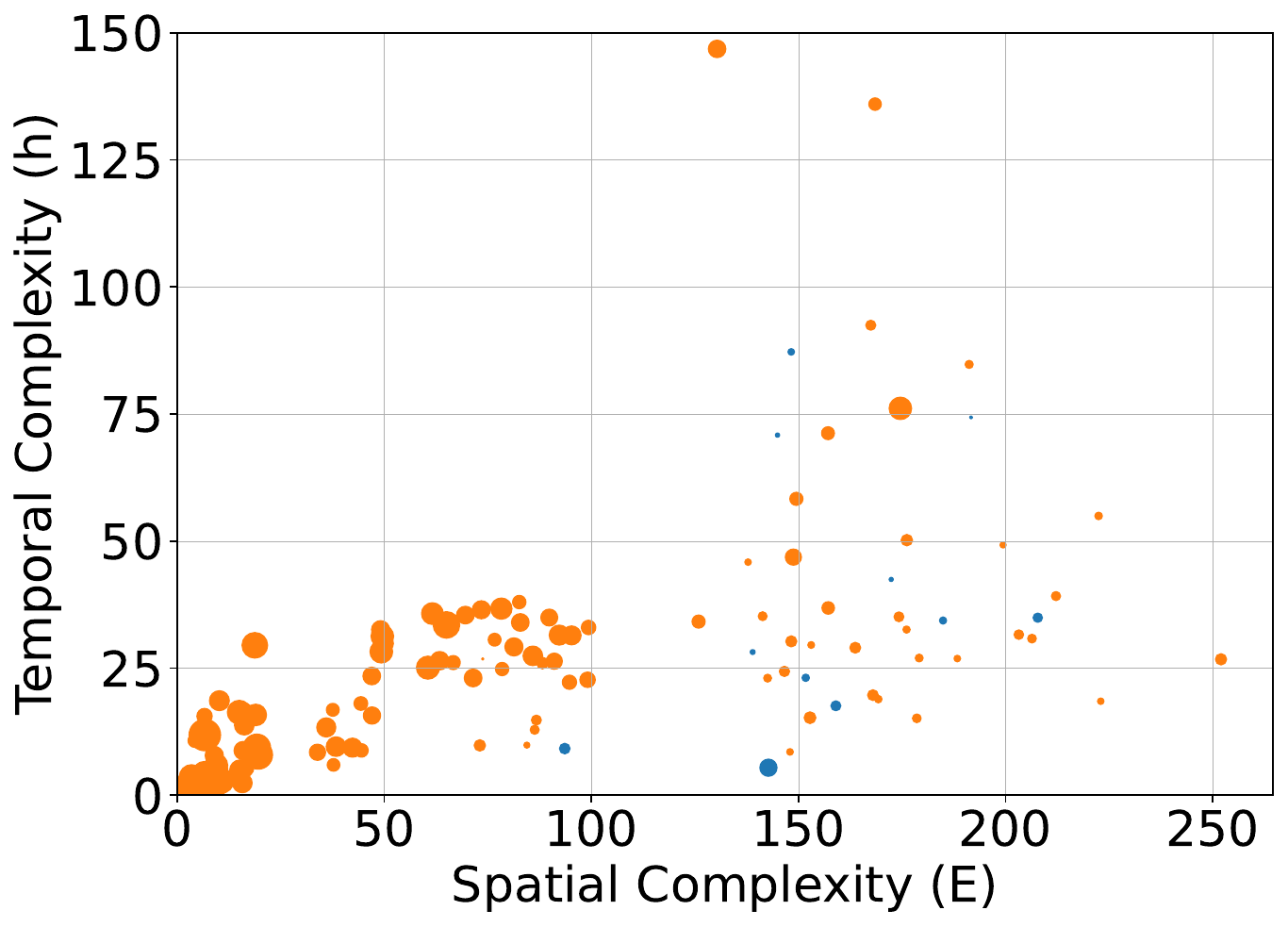}
    \caption{BDBR/BDT of Inter4k videos of various complexities.}
    \label{spatial-temporal}
\end{figure}

\begin{table*}[!t]
\small
\centering
\caption{The performance of MEVHAS and the fast preset compared to the medium preset.}
\resizebox{0.5\linewidth}{!}{
\label{table_BDPSNR_BDBR}
\begin{tabular}{|c|c|c|c|}
\hline
\textbf{Preset} &
  \textbf{BDT (\%)} &
  \textbf{BDBR (\%)} &
  \textbf{BDBR/BDT} \\ \hline
\textbf{MEVHAS} &
  16.73 &
  2.11 &
  0.12 \\ \hline
\textbf{Fast} &
  60.44 &
  14.98 &
  0.25 \\ \hline
\end{tabular}}
\end{table*}

\begin{figure*}[ht]
    \centering
    \captionsetup{justification=centering}
    \subfloat[]{
        \includegraphics[width=0.35\linewidth]{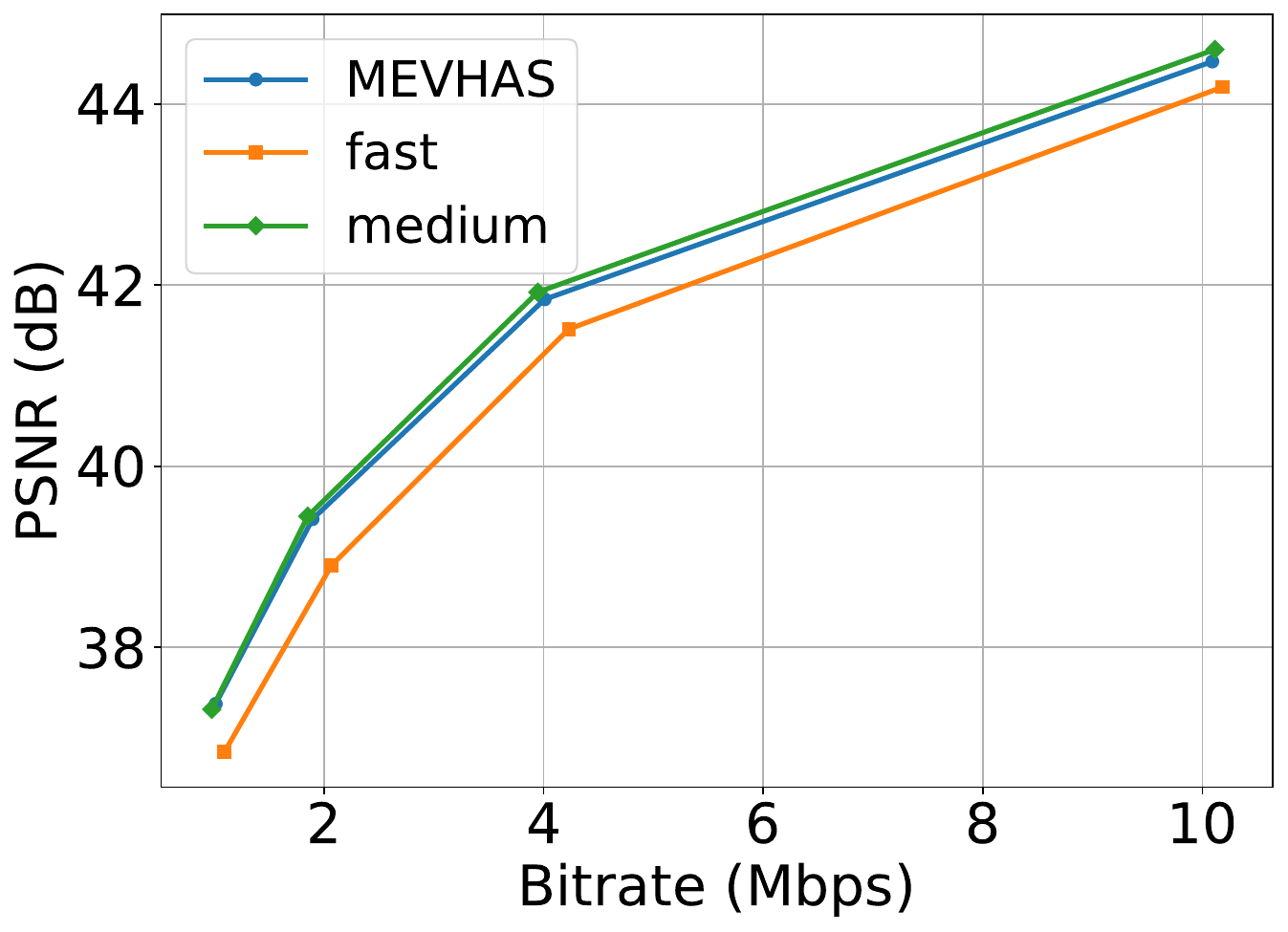}
        \label{seq-brVsPsnr}
    }
    \subfloat[]{
        \includegraphics[width=0.35\linewidth]{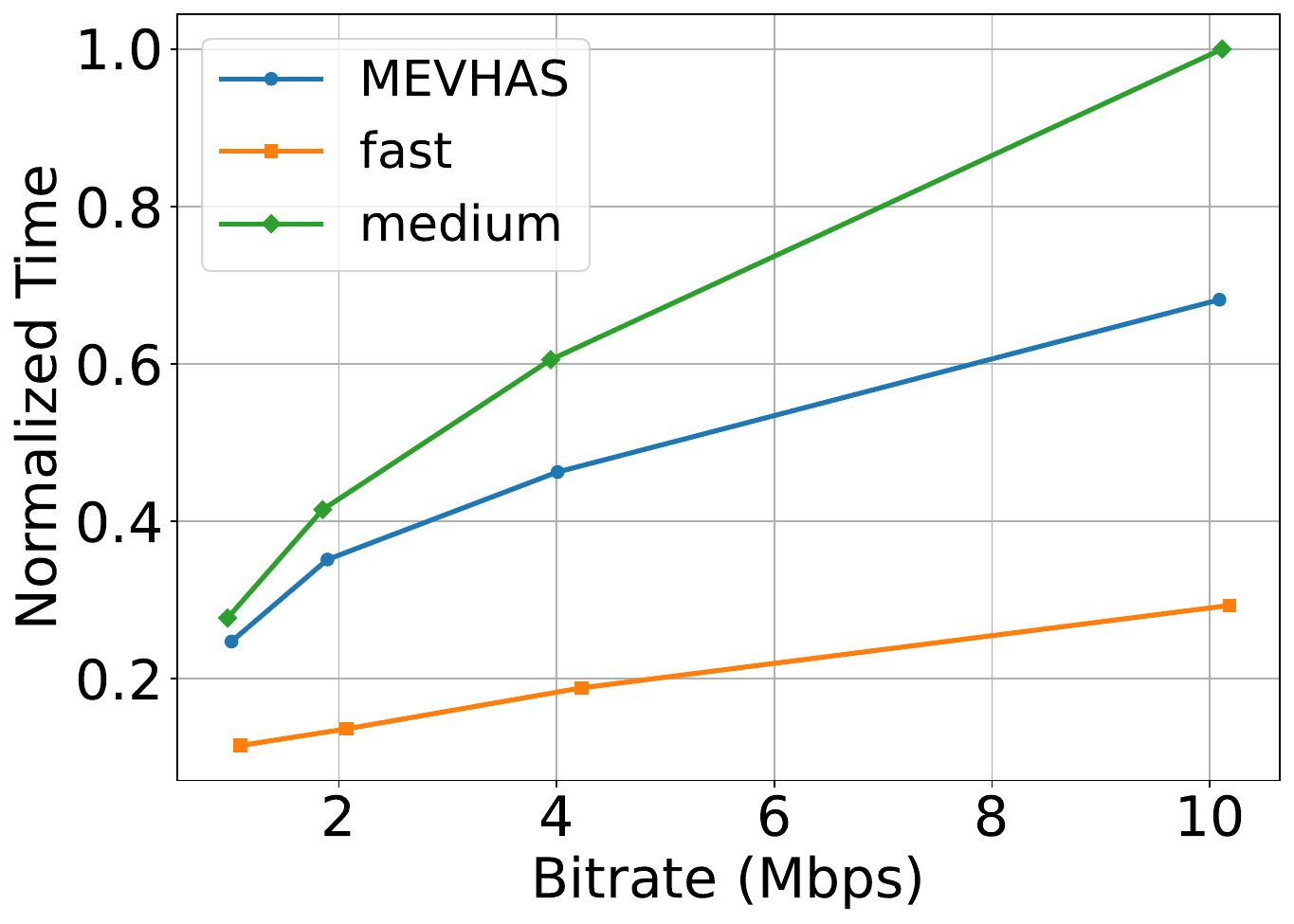}
        \label{seq-brVsTime}
    }
    \caption{(a) PSNR vs. bitrate and (b) Time vs. bitrate of fast, medium, and proposed MEVHAS method.}
    \label{fig:bitrate_psnr_time}
\vspace{-1em}
\end{figure*}
\section{Experimental Results}

\subsection{Setup}
The results are evaluated on 130 random Inter4K videos containing different complexities~\cite{stergiou2022adapool}. The complexity of the videos is determined using VCA \cite{menon2022vca} that extracts the spatial (E) and temporal (h) complexities of the videos.  
Each video is encoded on a server with an Intel Xeon Gold 5220R CPU @2.2 GHz using a single thread across four QPs: $27$, $32$, $37$, and $42$. 
Encodings are performed using standalone VVenC ver-1.7.0-rc with medium and fast presets using random access configuration and a Group of Pictures (GOP) of $32$. 
The proposed method is based on the medium preset, employing the same configuration.
To quantitatively evaluate MEVHAS, Bjøntegaard Delta Bit Rate (BDBR) using Peak Signal-to-Noise Ratio (PSNR) is used as the quality measure. 
Additionally, for time-saving, the research uses Bjøntegaard Delta Time Saving (BDT) \cite{herglotz2024energy}, which is the amount of time saved when the quality of the videos is the same. 
BDBR/BDT, which integrates both compression efficiency and computational performance, comprehensively assesses the bitrate reduction achieved relative to the time taken. Lower values signify enhanced overall performance.

\subsection{Discussion and Results}



This section highlights the improvements of MEVHAS over VVenC presets regarding both rate-distortion performance and encoding efficiency (BDBR/BDT).   

To analyze all videos, the results are shown in Fig.~\ref{spatial-temporal}.
The x-axis represents the spatial complexity (E) of a video, while the y-axis indicates the temporal complexity of a video. The dot sizes reflect the degree of improvement in BDBR/BDT compared to the fast preset, with larger dots indicating greater enhancements. Blue dots identify cases where the MEVHAS performs worse than the fast preset in terms of BDBR/BDT, highlighting trade-offs in specific video scenarios.



Table~\ref{table_BDPSNR_BDBR} presents the average results across all randomly selected Inter4K videos. 
It is evident that the MEVHAS achieves an average encoding time reduction of approximately 17\%. Notably, for the same amount of time saved, MEVHAS demonstrates superior BDBR/BDT performance of $0.12$ compared to the fast preset with a BDBR/BDT of $0.25$, indicating a more favorable trade-off between compression efficiency and encoding speed.

Figure~\ref{seq-brVsPsnr} illustrates the compression efficiency of the medium preset, fast preset, and MEVHAS for a representative Inter4K video (0402). The results show that the MEVHAS compression efficiency is closer to that of the medium preset, outperforming the fast preset. Additionally, Figure~\ref{seq-brVsTime} depicts the encoding time reduction for the same Inter4K sequence (0402). It can be seen that the MEVHAS saves more time at higher bitrates than at lower bitrates. This is because, although partitioning patterns are more similar to those at lower bitrates, the encoding speed for lower bitrates is already much faster~\cite{liu2022statistical}, leaving less potential for further time savings.




\section{Conclusions}

This paper introduces a fast and efficient multi-resolution encoding approach for HTTP adaptive streaming, significantly optimizing the rate-distortion search process. By leveraging an intermediate quality representation from a lower resolution as a reference, the encoding time for all higher resolution dependent representations is substantially reduced. The proposed method, MEVHAS, demonstrates outstanding performance, achieving a superior BDBR/BDT of 0.12 compared to the widely used fast preset, which yields a BDBR/BDT of 0.25 across videos of varying complexity. Additionally, MEVHAS cuts encoding time by an average of 17\% compared to the popular medium preset, offering a powerful solution for rapid and high-quality adaptive streaming.  

\section*{Acknowledgment}
The financial support of the Austrian Federal Ministry for Digital and Economic Affairs, the National Foundation for Research, Technology and Development, and the Christian Doppler Research Association, is gratefully acknowledged. Christian Doppler Laboratory ATHENA: \url{https://athena.itec.aau.at/}.

\vspace{12pt}
\balance
\bibliographystyle{ieeetr}
\bibliography{bibliography.bib,references_hadi}

\end{document}